%% file: main.tex
\documentclass[conference]{IEEEtran}
\usepackage[T1]{fontenc}
\usepackage[utf8]{inputenc}
\usepackage{cite}
\usepackage{graphicx}
\usepackage{amsmath,amssymb}
\usepackage{booktabs}
\usepackage{hyperref}
\usepackage{xcolor}
\usepackage{listings}
\usepackage{ifthen}
\usepackage[linesnumbered,ruled,vlined]{algorithm2e}

\usepackage{color}
\usepackage{xcolor}
\usepackage{tcolorbox}
\usepackage{listings}
\usepackage{enumitem}

\usepackage{threeparttable}

\lstdefinestyle{promptstyle}{
    basicstyle=\fontsize{6}{8.4}\selectfont\ttfamily,
    breaklines=true,
    columns=flexible,
    backgroundcolor=\color{blue!5},
    frame=single,
    frameround=tttt,
    rulecolor=\color{blue!30},
    commentstyle=\color{black},
    numberstyle=\fontsize{6}{8.4}\selectfont\color{black},
    showstringspaces=false,
    keywordstyle=,
    morekeywords={},
    moredelim=[is][\color{blue}\bfseries]{\{}{\}},
    moredelim=[s][\color{gray}\bfseries]{IMPORTANT:}{present.},
    moredelim=[s][\color{gray}\bfseries]{DEFINITIONS:}{wrong},
    moredelim=[s][\color{gray}\bfseries]{Provide}{fields:},
    moredelim=[s][\color{green!30!black}]{[}{]},
    moredelim=[s][\color{green!30!black}]{\{}{\}},
}

\lstdefinelanguage{Swift}{
  keywords={struct,func,class,import,let,var,if,else,while,for,return,guard,extension,enum,protocol,public,private,internal,open,static},
  sensitive=true,
  comment=[l]//,
  morecomment=[s]{/*}{*/},
  morestring=[b]"
}

\lstdefinelanguage{JSON}{
  morestring=[b]",
  showstringspaces=false,
  alsoletter=:_-\",
  basicstyle=\ttfamily\footnotesize,
}

\newboolean{showcomments}
\setboolean{showcomments}{true}
\ifthenelse{\boolean{showcomments}}
 { \newcommand{\mynote}[2]{
      \fbox{\bfseries\sffamily\scriptsize#1}      {\small$\blacktriangleright$\textsf{\emph{#2}}$\blacktriangleleft$}}}
        { \newcommand{\mynote}[2]{}}

\begin{document}

\title{LLMShot: Reducing snapshot testing maintenance via LLMs}
\author{
  \IEEEauthorblockN{Ergün Batuhan Kaynak}
  \IEEEauthorblockA{Bilkent University \\ Turkey \\ batuhan.kaynak@bilkent.edu.tr}
\and
  \IEEEauthorblockN{Mayasah Lami}
  \IEEEauthorblockA{Bilkent University \\ Turkey \\ m.lami@bilkent.edu.tr}
\and
  \IEEEauthorblockN{Sahand Moslemi}
  \IEEEauthorblockA{Bilkent University \\ Turkey \\ sahand.moslemi@bilkent.edu.tr	}
\and
  \IEEEauthorblockN{Anil Koyuncu}
  \IEEEauthorblockA{Bilkent University \\ Turkey \\ anil.koyuncu@cs.bilkent.edu.tr}
}

\maketitle

\input{abstract}

\IEEEpeerreviewmaketitle

\input{introduction}

\input{background}
\input{methodology}

\input{setup}

\input{evaluation}

\input{discussion}

\input{conclusion}

\bibliographystyle{IEEEtran}
\bibliography{main.bib}


\end{document}

%% file: abstract.tex
\begin{abstract}
Snapshot testing has emerged as a critical technique for UI validation in modern software development, yet it suffers from substantial maintenance overhead due to frequent UI changes causing test failures that require manual inspection to distinguish between genuine regressions and intentional design changes. This manual triage process becomes increasingly burdensome as applications evolve, creating a need for automated analysis solutions. This paper introduces LLMShot, a novel framework that leverages Vision-Language Models (VLMs) to automatically analyze snapshot test failures through semantic classification of UI changes. To evaluate LLMShot's effectiveness, we developed a comprehensive dataset using a feature-rich iOS application with configurable feature flags, creating realistic scenarios that produce authentic snapshot differences representative of real development workflows. Our evaluation using Gemma3 models demonstrates strong classification performance, with the 12B variant achieving over 84\% recall in identifying failure root causes while the 4B model offers practical deployment advantages with acceptable performance for continuous integration environments. However, our exploration of selective ignore mechanisms revealed significant limitations in current prompting-based approaches for controllable visual reasoning. LLMShot represents the first automated approach to semantic snapshot test analysis, offering developers structured insights that can substantially reduce manual triage effort and advance toward more intelligent UI testing paradigms.
\end{abstract}

\begin{IEEEkeywords}
snapshot testing, software testing, large language models, vision-language models, user interface testing, automated test analysis
\end{IEEEkeywords}


%% file: introduction.tex
\section{Introduction}
\label{sec:introduction}

Modern software development cycles demand rapid iteration and frequent code changes, creating significant challenges for development teams who must balance software quality while keeping development cycles short~\cite{ferrario2022applying}. Over the years, many tools and techniques have been developed to help engineers test software systematically at various levels.

Snapshot testing offers a practical approach for user interface (UI) components by capturing the current output of an application and comparing it against future runs to detect changes~\cite{emp_study_snapshot, cruz2023snapshot}. Unlike unit and functional tests that verify correctness, snapshot tests simply check whether anything has changed from a previous known state, making them particularly useful for catching unintended modifications during development. Currently used in several large software companies~\cite{Hanlon_2022}, snapshot testing is adopted due to its ease of implementation, allowing developers to write more tests and consequently prevent regressions.

\begin{figure}
    \centering
    \includegraphics[width=0.9\linewidth]{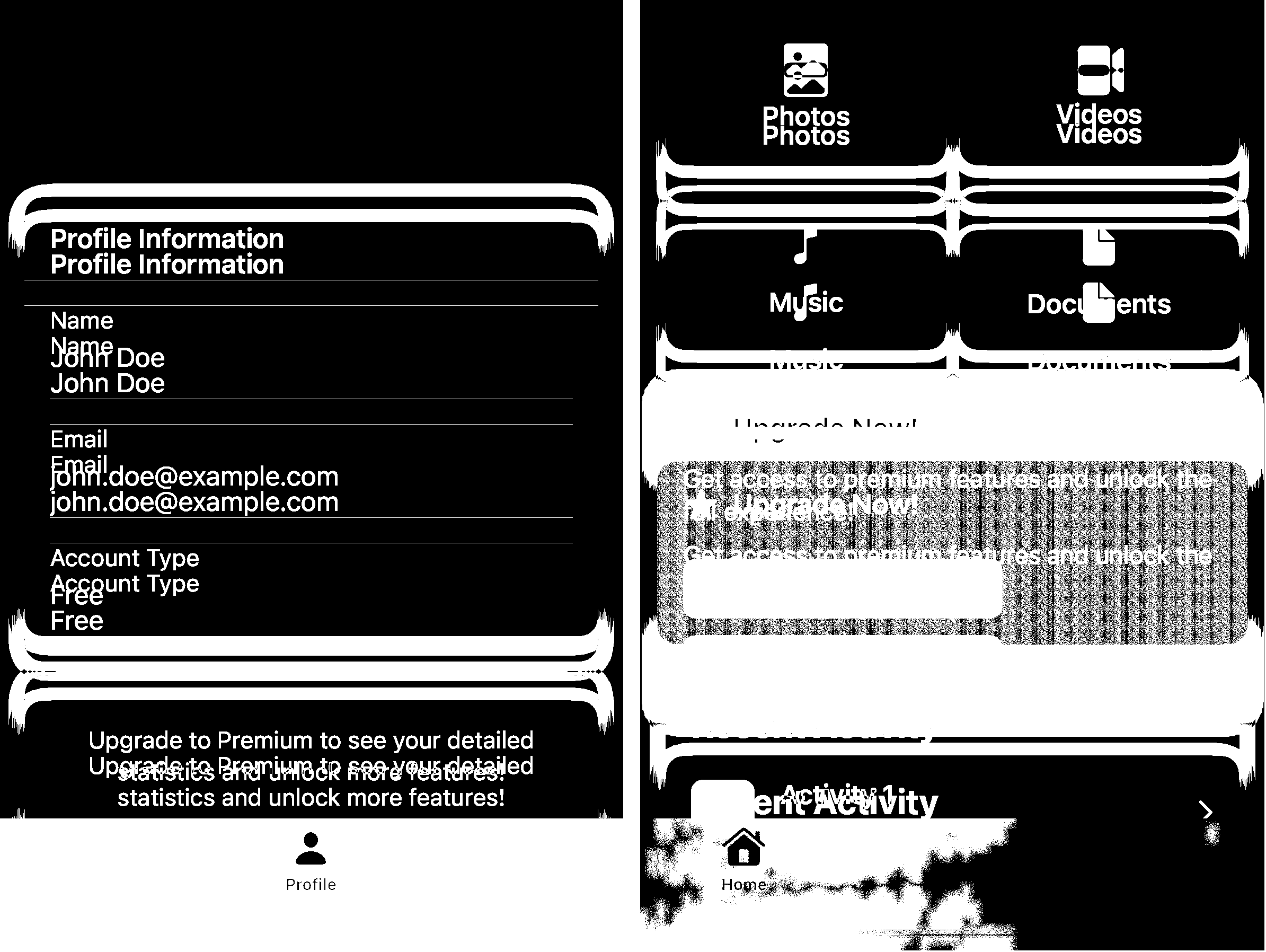}
    \caption{Two Example Diffs for the \texttt{PADDING\_CHANGE} Test Failure Class}
    \label{fig:padding_snapshot}
\end{figure}

Despite its benefits, snapshot testing faces significant maintenance challenges. Like other GUI testing techniques~\cite{sapienz16, stoat18, combodroid19}, current approaches frequently fail to comprehend semantic information~\cite{gptdroid24, su2021benchmarking}. Snapshot tests focus solely on identifying visual differences without providing actionable insights, instead only indicating that changes have occurred without meaningful context. When failures occur, as illustrated in Figure~\ref{fig:padding_snapshot}, developers must triage each failure to determine whether it represents a regression or reflects intentional changes. Current snapshot testing practices suffer from two key limitations: (1) they do not provide detailed explanations of failures, and (2) they produce frequent false positives from minor, acceptable changes~\cite{cruz2023snapshot}.

While recent research explored smarter visual oracles through tools like DiffDroid~\cite{diffdroid20}, Owl Eyes~\cite{owleyes20}, and GLIB~\cite{glib21} that employ deep learning classifiers for visual defects, these approaches fail to contextualize the changes. Existing guidelines recommend detailed descriptions for snapshot tests~\cite{cruz2023snapshot}, but current tools provide no automated support for generating such explanations.

To address these challenges, we introduce LLMShot, a framework that leverages VLMs to automate snapshot test failure analysis. LLMShot employs Gemma3 models to categorize, quantify, and explain visual discrepancies through semantic classification into a predefined taxonomy. Unlike existing pass/fail approaches, our framework provides semantic understanding of UI changes, distinguishing between genuine regressions and intentional design modifications.


Our evaluation demonstrates strong classification performance, with the Gemma3 12B model achieving 84.21\% recall. The more efficient 4B model also performs robustly, reaching 78.95\% recall. However, our exploration of the ignore mechanism reveals that current prompting techniques are insufficient for complex controllable reasoning, highlighting a key challenge for future research.

\noindent The main contributions of this work are:

\begin{itemize}[leftmargin=*]
\item The design and implementation of LLMShot, the first automated framework for the semantic analysis of snapshot test failures. LLMShot categorizes visual discrepancies according to a structured taxonomy of change types and through its structured JSON output, provides a pipeline suitable for CI/CD integration.

\item A new, publicly available dataset\footnote{\url{https://zenodo.org/records/15876539}} for snapshot testing research. Our dataset provides ground truth labels based on our UI failure taxonomy and contains realistic and challenging UI changes generated from a feature-rich iOS application.
\end{itemize}

%% file: background.tex
\section{Background and Related Work}
\label{sec:related_work}

\subsection{Visual GUI and Snapshot Testing}
Early automated GUI testing aimed to maximize functional coverage. This foundational work includes tools for event-sequence generation and exploration like Sapienz~\cite{sapienz16} and Stoat~\cite{stoat18}, efficiency improvements through techniques like combinatorial reduction~\cite{mirzaei16}, and automated test script generation from recordings or traces~\cite{combodroid19, V2S, CAPdroid}. More recently, reinforcement learning has been applied to improve navigation and bug detection~\cite{webexplor23, qexplore24}. While comprehensive, this research primarily targets functional correctness, leaving the distinct challenge of non-functional visual regressions largely unaddressed.

Snapshot testing has gained popularity precisely to fill this gap by comparing rendered UI states against baselines. However, it remains largely understudied in the literature. Existing studies are often preliminary; for instance, Bui and Rocha~\cite{bui2023snapshot} compiled a dataset of static project attributes without executing tests, while a grey literature review by Cruz et al.~\cite{cruz2023snapshot} noted the technique's benefits but also its fragility and high rate of false positives.

In parallel with the rise of snapshot testing, broader research has explored smarter visual oracles to address the general fragility of visual UI validation. These include tools for detecting rendering inconsistencies across devices like DiffDroid~\cite{diffdroid20}, layout issues using metamorphic testing~\cite{dvermin21}, and visual glitches via deep learning classifiers~\cite{owleyes20, nighthawk22, glib21}. However, a critical limitation unites these advanced oracles: while they excel at \textit{detecting} visual defects, they do not \textit{explain} them. They fail to provide the semantic context needed to determine if a failure is a genuine regression or an intentional change.

\subsection{LLM‑based Semantic Analysis for UI Verification}
The advent of Vision-Language Models (VLMs) has unlocked new potential for semantic analysis in UI testing. Early successes like GPTDroid~\cite{gptdroid24} and VisionDroid~\cite{visiondroid25} have already leveraged VLMs to find complex functional bugs and improve test generation by reasoning about on-screen content.

Despite these advances, their application has been concentrated on functional bug detection, leaving the core problem of snapshot test analysis unexplored. The primary challenge in snapshot testing is not detection, but the high manual effort required to triage failures caused by intentional UI changes~\cite{emp_study_snapshot}. While guidelines recommend writing tests with ``detailed and clear descriptions''~\cite{cruz2023snapshot}, no tools provide this automatically. LLMShot is the first work to bridge this explanatory gap. It applies VLMs to semanticize the validation process itself, providing the contextual analysis needed to distinguish intentional changes from true regressions.

%% file: methodology.tex
\section{Methodology}
\label{sec:methodology}


Our primary objective is to distinguish genuine UI regressions from intentional design changes by providing automated, explanatory analysis for failing snapshot tests. To this end, we introduce the LLMShot framework, which we evaluate on a new dataset designed specifically for this task.

\subsection{LLMShot}

LLMShot processes snapshot test failures using VLMs. The core of the framework is a structured prompt, shown in Listing~\ref{lst:prompttemplate}, which was developed through an iterative design process.
\input{prompt}

Following the instructions in this prompt, LLMShot operates on a specific set of inputs to produce a structured analysis. The primary \textbf{input} is a tuple of the three images referenced in the prompt (\texttt{before}, \texttt{after}, and \texttt{diff}). The corresponding \textbf{output} is the single JSON object specified in the prompt.

The JSON output provides a multi-faceted analysis. The \textit{Change Classification} assigns one or more labels from the predefined categories (Table~\ref{tab:snapshot_failure_reasons}). A key feature is that if the model identifies a change that does not fit the existing taxonomy, it can dynamically create a new category by giving a descriptive name to the newly identified change type. The \textit{Quantitative Measurements} include a pixel-wise difference score (on a 0-1 scale where 0 is identical) and a semantic difference score (on a 0-1 scale where 0 means all expected components are correct). Finally, the output provides a list of specific \textit{Affected UI Elements} and a concise \textit{Natural Language Explanation} analyzing the key differences observed between the snapshots.

\subsection{Extension: Ignore Category Mechanism}
In practice, developers often need to ignore expected changes (e.g., a planned color update) to focus on unexpected ones that may be genuine regressions. To simulate this critical workflow, we developed an ``ignore mechanism'' extension. The implementation modifies the core prompt by appending an instruction for the VLM to disregard a specified category while maintaining its sensitivity to any other modifications. The template for this extension is shown in Listing~\ref{lst:ignoretemplate}.




\input{ignore_prompt}

\input{dataset}

%% file: prompt.tex
\begin{figure}[t]
\centering
\begin{lstlisting}[style=promptstyle,caption={LLM prompt framework of LLMShot},label={lst:prompttemplate}]
Compare three images:
1. REFERENCE: {reference} --- correct baseline image
2. FAILURE: {failure} --- current test image
3. DIFF: {difference} --- pixel-level diff image

Analyze the failure by examining ALL three images carefully.

POSSIBLE CATEGORIES: {categories}
Select ALL applicable categories. Create new ones as "UNKNOWN-<reason>".

DEFINITIONS:
- Pixel-wise diff (0-1): 0 = identical, 1 = completely different
- Semantic diff (0-1): 0 = all components present/correct, 1 = all missing/wrong

Respond with valid JSON only:
  "reasons": ["category1", "UNKNOWN-X"],
  "pixel_diff": 0.XX,
  "semantic_diff": 0.XX,
  "elements": ["specific UI component names"],
  "explanation": "Clear, concise analysis of key differences"
\end{lstlisting}
\end{figure}

%% file: ignore_prompt.tex



\begin{lstlisting}[style=promptstyle,caption={Ignore reason template for LLM},label={lst:ignoretemplate}]
{core_prompt}

IGNORE the following aspect of the differences: {ignore_reason}
This difference is acceptable; focus on other differences that might exist.
\end{lstlisting}

%% file: dataset.tex
\subsection{Evaluation Dataset}

A key goal in snapshot testing is developing semantic understanding to distinguish intentional feature updates from genuine regressions. This requires realistic test scenarios with meaningful labels that go beyond a binary pass/fail result. The only prior snapshot-specific dataset, from Bui et al.~\cite{bui2023snapshot} provides a valuable foundation of static project analysis, but lacks the executable tests and visual snapshot content essential for analyzing failure root causes.

To fill this critical gap, we introduce a new, publicly available dataset designed specifically for this task. We generate this dataset from a purpose-built iOS application, using its integrated feature flags to produce authentic UI changes that mimic common development scenarios. Each resulting failure instance is then manually classified by two authors and labeled according to our detailed taxonomy of UI changes (Table~\ref{tab:snapshot_failure_reasons}). This taxonomy covers a wide range of modifications, including visual changes (e.g., color, padding), content updates (e.g., text, data), structural modifications (e.g., layout), and behavioral changes (e.g., animations).

This process results in a novel dataset containing not only baseline and failure snapshots but also rich contextual metadata and corresponding ground truth labels. To the best of our knowledge, it is the first publicly available dataset to provide these dynamic artifacts, thereby enabling new opportunities for the rigorous, empirical evaluation of tools for semantic snapshot test analysis.

\input{categories_table}

%% file: categories_table.tex
\begin{table}[!htb]
\centering
\scriptsize
\caption{Snapshot Test Failure Categories and Descriptions}
\label{tab:snapshot_failure_reasons}
\renewcommand{\arraystretch}{0.9}
\begin{tabular}{@{}p{2.2cm}p{6cm}@{}}
\textbf{Category} & \textbf{Description} \\
\midrule
\texttt{COLOR\_CHANGE} & Different color values due to styling updates. \\
\midrule
\texttt{PADDING\_CHANGE} & Change in margins/spacing; layout shifts. \\
\midrule
\texttt{CONTENT\_CHANGE} & Different content (e.g. image) with changed meaning. \\
\midrule
\texttt{LAYOUT\_CHANGE} & Components repositioned/resized, changing structure. \\
\midrule
\texttt{TEXT\_CHANGE} & Text string changed, altering displayed message meaning. \\
\midrule
\texttt{ANIMATION\_PHASE} & Snapshot taken mid-animation, showing intermediate state. \\
\midrule
\texttt{ANIMATION\_CHANGE} & Change in animation duration/timing alters output. \\
\midrule
\texttt{SEMANTIC\_CHANGE} & Behavior changed (toggle on/off) without layout/text change. \\
\midrule
\texttt{UNKNOWN\_<T>} & Change not from above categories; \texttt{T} names new reason. \\
\end{tabular}
\end{table}

%% file: setup.tex
\section{Setup}
\label{sec:setup}

To evaluate LLMShot, we detail the construction of our dataset, our model configuration, and the metrics used for assessment.

\subsection{Dataset Construction}

We generate our dataset from a purpose-built, feature-rich iOS application, which we develop in SwiftUI specifically for this study. We design the app with a non-trivial architecture (MVVM, tab-based navigation) and dynamic UI elements that change based on user state and active feature flags to provide a realistic evaluation context.

\textbf{Data Generation.} We use the application's 19 runtime-configurable feature flags, a standard industry practice that allows for the controlled creation of diverse UI modifications such as changes to content visibility, layout, and dynamic animations. The combination of these flags with different user states (free vs. premium) creates diverse scenarios where the same flag can produce different visual outcomes depending on the UI context. We execute each of our 17 test cases using standard XCTest and XCUITest suites, with extensions to support scrolling and element interactions. For capturing the images themselves, we use the \texttt{swift-snapshot-testing} library~\cite{Pointfreeco}, running each test twice: once with default settings to capture a baseline, and once with a feature flag enabled to produce a failure. Two tests involve multi-flag activations to create more complex scenarios. The test execution environment is Xcode 16.3 on an iPhone 15 Pro running iOS 17.2.

\textbf{Ground Truth Labeling.} To establish a reliable ground truth, two of the authors independently classify each resulting failure according to the categories in Table~\ref{tab:snapshot_failure_reasons}. They then resolve all disagreements through a joint discussion to finalize the labels.

\textbf{Dataset Characteristics.} The distribution of flag categories across the 17 tests shows that \texttt{COLOR\_CHANGE} is the most prevalent type of UI discrepancy, occurring in 5 tests. Three categories—\texttt{PADDING\_CHANGE}, \texttt{CONTENT\_CHANGE}, and \texttt{LAYOUT\_CHANGE}—each appear in 3 tests. \texttt{TEXT\_CHANGE} and \texttt{ANIMATION\_PHASE} each occur in 2 tests, while \texttt{ANIMATION\_CHANGE} is the least frequent, appearing in only 1 test.

The dataset exhibits a mean pixel difference of 0.053 $\pm$ 0.054 (mean $\pm$ std) across all test cases, calculated using Equation~\ref{eq:pixel_diff}. This indicates that, on average, the reference and failure images differ by approximately 5.3\% at the pixel level. The substantial variation across test cases demonstrates dataset diversity. We quantify image-level differences using a normalized pixel-wise difference score:

{\small
\begin{equation}
\label{eq:pixel_diff}
D_p = \frac{1}{N} \sum_{i=1}^{N} \frac{|R_i - F_i|}{255}
\end{equation}
}

where $N$ is the number of pixels, $R_i$ and $F_i$ are RGB values at pixel $i$ in reference and failure images, respectively.

\subsection{Model Configuration}
We select the Gemma3~\cite{gemma3} family of models for our evaluation for two key reasons. First, they are state-of-the-art, open-weights models that support multi-image prompts, a critical feature for our framework which requires analyzing reference, failure, and diff images simultaneously. Second, their availability in multiple sizes allows us to analyze performance-to-scale trade-offs. We evaluate LLMShot using the 4B and 12B parameter variants with a temperature of 0.1 through the Ollama framework~\cite{ollama2025}. The prompt framework (Listing~\ref{lst:prompttemplate}) guides the LLM to analyze reference, failure, and diff images systematically, producing structured JSON responses. 


\subsection{Evaluation Metrics}
Performance assessment combines classification and difference estimation metrics.

Classification metrics include \textit{Test Case Hit Rate} (percentage of cases with at least one correct ground truth category), \textit{Recall} (proportion of ground truth labels correctly predicted), \textit{Precision} (proportion of correct model predictions), \textit{F1-Score} (balanced precision-recall measure penalizing false positives), \textit{Average Labels Per Test} (model's prediction tendency), and \textit{Unknown Rate} (cases defaulting to \texttt{UNKNOWN\_<T>} classifications).

For visual discrepancy quantification, \textit{Ground Truth Pixel Difference} uses deterministically calculated values (Equation~\ref{eq:pixel_diff}), \textit{Predicted Pixel Difference} represents LLM estimates, \textit{Error Pixel Difference} measures absolute prediction errors calculated as $|D_p^{predicted} - D_p^{ground\_truth}|$, and \textit{Semantic Difference} captures model-generated perceptual dissimilarity scores.

For ignore mechanism evaluation, \textit{Ignore Compliance Rate} measures the percentage of tests where the specified ignore category was not predicted (indicating successful adherence to ignore instructions), and classification metrics are calculated after removing ignored categories from both predictions and ground truth to assess performance on remaining categories.

%% file: evaluation.tex
\section{Evaluation}
\label{sec:results}

\subsection{RQ1: How does LLMShot perform in Failure Classification?}

The primary objective is to establish whether visual language models can accurately categorize different types of UI changes in snapshot test failures. To evaluate this capability, we applied the LLMShot prompt framework to our dataset using both Gemma3-4B and Gemma3-12B models.

Table~\ref{tab:rq13} shows LLMShot's classification performance across model sizes. The larger Gemma3-12B model outperformed the smaller Gemma3-4B model across all metrics, achieving over 80\% Hit Rate and 84.21\% Recall. However, Precision was notably lower for both models, indicating a tendency to generate some incorrect predictions alongside correct ones. F1-Scores of 74.42\% and 66.67\% respectively demonstrate balanced performance when accounting for these false positives. The superiority of the larger model is further evidenced by its lower `Unknown Rate` (5.88\% vs 11.76\%) and more accurate pixel difference estimation. Both models showed conservative prediction behavior, averaging fewer than 1.5 labels per test.

\textbf{Finding 1:} Our results indicate that LLMShot demonstrates strong potential in automatically classifying snapshot test failure root causes. The larger 12B model shows superior performance in both classification and quantitative reasoning, though both models exhibit a trade-off between high recall and moderate precision.




\input{rq13_table}

\subsection{RQ2: How does the ignore-category mechanism affect the performance of LLMShot?}

In real-world development, a critical task is to selectively ignore failures from intentional changes while still detecting regressions. We investigate whether our prompt-based ignore mechanism can achieve this, using the more efficient Gemma3:4b model. We evaluate two strategies: ``Ignore From Analysis'' (IFA), which instructs the model to ignore the category it first identified, and ``Ignore From Ground Truth'' (IFGT), which instructs it to ignore a known ground truth category.

The results, presented in Table~\ref{tab:ignore_modes_gemma3_4b}, reveal a significant performance degradation across both strategies. The IFA strategy performed worst, with its Hit Rate dropping to 18.75\% and its F1-Score decreasing sharply to 13.79\%. The IFGT strategy presented a more complex outcome. While its Recall reached 100.00\%, this result is misleading as it was coupled with an extremely low precision of 9.09\%. This suggests the model adopted a noisy prediction strategy, which is further reflected by its marginal F1-Score of only 16.67\%.

Furthermore, both strategies demonstrated poor instruction compliance, with Ignore Compliance Rates of only 31.25\% for IFA and 35.29\% for IFGT. This indicates that the model frequently failed to disregard the specified ignore category. This difficulty is also reflected in other metrics: pixel difference errors increased substantially (0.157$\pm$0.244 for IFA and 0.287$\pm$0.339 for IFGT vs 0.095$\pm$0.161 baseline), and the unknown classification rate for IFA rose to 25\%.

\textbf{Finding 2:} Current prompt-based ignore mechanisms struggle severely with selective change filtering. They exhibit poor instruction compliance and suffer a substantial degradation across all classification metrics. The high recall observed in one strategy is an artifact of its extremely low precision, rendering the approach unreliable for practical use.

\input{rq2_table}

%% file: rq13_table.tex
\begin{table}[h!]
\centering
\scriptsize
\setlength{\tabcolsep}{2pt} 
\renewcommand{\arraystretch}{1.05} 
\caption{Performance of LLMShot}
\label{tab:rq13}
\resizebox{0.7\linewidth}{!}{
\begin{tabular}{lcc}
\textbf{Model} & \textbf{gemma3:4b} & \textbf{gemma3:12b} \\
\midrule
Hit Rate (\%) & 76.47 & 82.35 \\
Recall (\%) & 78.95 & 84.21 \\
Precision (\%) & 57.69 & 66.67 \\
F1-Score (\%) & 66.67 & 74.42 \\
Avg. \# label/test & 1.53 $\pm$ 0.50 & 1.41 $\pm$ 0.49 \\
Unknown Rate (\%) & 11.76 & 5.88 \\
\midrule
Predicted & 0.126 $\pm$ 0.186 & 0.066 $\pm$ 0.058 \\
Error & 0.095 $\pm$ 0.161 & 0.059 $\pm$ 0.048 \\
\midrule
Semantic Diff. & 0.921 $\pm$ 0.046 & 0.079 $\pm$ 0.079 \\
\bottomrule
\end{tabular}
}
\vspace{-2mm} 
\begin{minipage}{0.8\linewidth}
\centering
\begin{tablenotes}
\item[]\textbf{Avg. \# label/test:} Average Number of Predicted Labels per Test
\item[]\textbf{Hit Rate:} Test Case Hit Rate, \textbf{Semantic Diff.:} Semantic Difference
\end{tablenotes}
\end{minipage}
\end{table}

%% file: rq2_table.tex

\begin{table}[!htb]
\centering
\scriptsize
\setlength{\tabcolsep}{2pt} 
\renewcommand{\arraystretch}{1.05} 
\caption{Performance of LLMShot$^{\text{gemma3:4b}}$ with Ignore Mechanism}
\label{tab:ignore_modes_gemma3_4b}
\resizebox{0.8\linewidth}{!}{
\begin{tabular}{@{}lccc@{}}
\toprule
\textbf{Metric} & \textbf{LLMShot} & \textbf{IFA} & \textbf{IFGT} \\
\midrule
Hit Rate (\%) & 76.47 & 18.75 & 41.18 \\
Recall (\%) & 78.95 & 28.57 & 100.00 \\
Precision (\%) & 57.69 & 9.09 & 9.09 \\
F1-Score (\%) & 66.67 & 13.79 & 16.67 \\
Avg. \# label/test & 1.53$\pm$0.50 & 1.38$\pm$0.60 & 1.29$\pm$0.96 \\
\midrule
Unknown Rate (\%) & 11.76 & 25.00 & 11.76 \\
IC Rate (\%)$^{\dagger}$ & --- & 31.25 & 35.29 \\
\midrule
Predicted & 0.126$\pm$0.186 & 0.194$\pm$0.255 & 0.332$\pm$0.326 \\
Error & 0.095$\pm$0.161 & 0.157$\pm$0.244 & 0.287$\pm$0.339 \\
\midrule
Semantic Diff. & 0.921$\pm$0.046 & 0.894$\pm$0.050 & 0.891$\pm$0.060 \\
\bottomrule
\end{tabular}
}
\vspace{-2mm} 
\begin{minipage}{0.9\linewidth}
\centering
\begin{tablenotes}
\item[]\textbf{Avg. \# label/test:} Average Number of Predicted Labels per Test
\item[]\textbf{IC:} Ignore Compliance
\end{tablenotes}
\end{minipage}
\end{table}

%% file: discussion.tex
\section{Discussion}
\label{sec:discussion}

Our evaluation demonstrates that LLMShot successfully addresses a fundamental gap in snapshot testing by providing semantic understanding of UI changes. Unlike existing tools that focus primarily on detection~\cite{diffdroid20, owleyes20}, LLMShot offers the explanatory analysis needed to distinguish between intentional design changes and genuine regressions. For practitioners, this suggests a path toward more intelligent CI/CD workflows where the manual triage of failures is significantly reduced, allowing developer effort to be focused on genuine issues. The strong classification performance indicates that VLMs possess sufficient visual reasoning capabilities for this practical task.

However, our exploration of ignore mechanisms reveals fundamental limitations in current prompting-based approaches. The substantial performance degradation when attempting to filter changes highlights a broader challenge for VLM-based systems: achieving fine-grained, controllable reasoning remains difficult. This finding points to a clear direction for future research: developing more robust methods beyond simple prompting, such as fine-tuning, is necessary to unlock the full potential of this technology.

Our dataset construction approach, using a purpose-built application with feature flags, demonstrates the importance of authentic test scenarios. To further advance this line of research, the community would benefit from larger, more diverse public datasets, ideally mined from real project histories. Additionally, future work should include user studies to qualitatively evaluate the impact of generated explanations on developer productivity.


\section{Threat to Validity}

\textbf{External Validity:} Our evaluation is conducted on a single iOS application, which limits the generalizability of our findings. The performance of LLMShot may differ on applications built with other frameworks (e.g., Android, web) or those with different design paradigms.

\textbf{Construct Validity:} Our predefined categories may not encompass all possible UI change types. We address this with the \texttt{UNKNOWN} label mechanism, but a broader study is needed. Furthermore, our author-labeled ground truth, while cross-validated between two authors, could benefit from additional expert validation in future work.

\textbf{Comparison to Baselines:} Our study does not include a head-to-head comparison with prior tools like DiffDroid~\cite{diffdroid20}. A direct comparison is challenging, as such tools are often platform-specific and operate on different principles (e.g., view hierarchy analysis) than our image-based approach, but establishing such baselines is an important direction for future work.

\textbf{Reliability:} LLM inherent randomness may produce varying outputs across runs. We mitigate this through low temperature settings and consistent hyperparameters, ensuring reproducible results.

%% file: conclusion.tex
\section{Conclusion}
\label{sec:conclusion}

This paper introduced LLMShot, the first framework designed to apply VLMs to the semantic analysis of snapshot test failures. Our evaluation demonstrates that LLMShot can accurately classify the root causes of UI changes with high recall, successfully moving beyond the simple pass/fail paradigm of traditional visual oracles.

The key contributions of this work are twofold: the LLMShot framework itself, which provides a novel approach for automated failure analysis, and a new, publicly available dataset with validated ground truth labels to enable further research in this area. Our empirical results highlight both the significant potential of VLMs for this task and the key challenges that remain, particularly in achieving fine-grained, controllable reasoning through prompting alone.

Ultimately, LLMShot establishes a foundation for a new class of intelligent UI testing tools. By translating ambiguous visual differences into structured, actionable insights, our work highlights a promising path toward reducing developer maintenance effort. It represents a critical first step in making snapshot testing a more scalable, sustainable, and insightful practice in modern software development.